\documentclass[conference]{IEEEtran}
\IEEEoverridecommandlockouts
\usepackage[table,dvipsnames]{xcolor}
\usepackage{csquotes}
\usepackage{acro}
\usepackage{array}
\usepackage{float}
\usepackage{tcolorbox}
\usepackage{subcaption}
\usepackage{booktabs}
\usepackage{tabularx}
\usepackage{multicol}
\usepackage{multirow}
\usepackage{makecell}
\usepackage{bm}
\usepackage{mathtools}
\usepackage[inline]{enumitem}
\usepackage{wasysym}
\usepackage{tikz}
\usepackage{amsmath}
\usepackage{arydshln}

\newtcolorbox[auto counter]{prompt}[2][]{%
colback=white,colframe=blue!20!white,coltitle=black, title=Prompt~\thetcbcounter: #2,#1}

\newtcolorbox[auto counter]{instruc}[2][]{%
colback=white,colframe=blue!50!white,coltitle=black, title=System Instruction~\thetcbcounter: #2,
fontupper=\small, before skip=4pt, after skip=4pt, top=3pt, bottom=3pt,#1}

\usepackage[nomessages]{fp}

\newlength{\maxlen}

\setlist{leftmargin=4mm}
\DeclareMathOperator*{\argmax}{argmax}

\newcommand{\inc}[1]{\textcolor{blue}{\tiny $\blacktriangle$#1}}
\newcommand{\dec}[1]{\textcolor{red}{\tiny $\blacktriangledown$#1}}

\DeclareAcronym{llm}{
    short = LLM,
    long = large language model
}

\DeclareAcronym{ohlc}{
    short = OHLC,
    long = open\, high\, low\, and close
}

\DeclareAcronym{mcc}{
    short = MCC,
    long = Matthews Correlation Coefficient
}

\DeclareAcronym{capm}{
    short = CAPM,
    long = Capital Asset Pricing Model 
}

\DeclareAcronym{hml}{
    short = HML,
    long = High-Minus-Low
}

\DeclareAcronym{mas}{
    short = MAS,
    long = multi-agent system
}

\DeclareAcronym{cot}{
    short = CoT,
    long = chain-of-thought
}

\DeclareAcronym{rag}{
    short = RAG,
    long = retrieval-augmented generation
}

\DeclareAcronym{sa}{
    short = SA,
    long = single agent
}

\DeclareAcronym{json}{
    short = JSON,
    long = JavaScript Object Notation
}
\usepackage{cite}
\usepackage{amssymb,amsfonts}
\usepackage{algorithmic}
\usepackage{graphicx}
\usepackage{textcomp}
\usepackage[colorlinks=false,pdfborder={0 0 0}]{hyperref}
\def\BibTeX{{\rm B\kern-.05em{\sc i\kern-.025em b}\kern-.08em
    T\kern-.1667em\lower.7ex\hbox{E}\kern-.125emX}}
\begin{document}

\title{LLM-Powered Multi-Agent System for Automated Crypto Portfolio Management}

\author{
\IEEEauthorblockN{Yichen Luo\IEEEauthorrefmark{1},
Yebo Feng\IEEEauthorrefmark{2}$^{*}$,
Jiahua Xu\IEEEauthorrefmark{1}\IEEEauthorrefmark{3},
Paolo Tasca\IEEEauthorrefmark{1}\IEEEauthorrefmark{3},
Yang Liu\IEEEauthorrefmark{2}}
\IEEEauthorblockA{\IEEEauthorrefmark{1}\textit{University College London}, London, United Kingdom\\
\{yichen.luo.22, jiahua.xu, p.tasca\}@ucl.ac.uk}
\IEEEauthorblockA{\IEEEauthorrefmark{2}\textit{Nanyang Technological University}, Singapore\\
\{yebo.feng, yangliu\}@ntu.edu.sg}
\IEEEauthorblockA{\IEEEauthorrefmark{3}\textit{Exponential Science}, London, United Kingdom}
}

\maketitle

\begin{abstract}
Cryptocurrency portfolio management requires the fusion of heterogeneous multi-modal signals — structured price and on-chain time series, unstructured news text, and computed technical indicators — under high-volatility, real-time constraints. While deep learning approaches demonstrate predictive capability, their opacity limits practical adoption, and single \ac{llm} agents struggle to process the breadth of modality-specific inputs required for robust decision-making. We propose a \ac{mas} framework in which three modality-specialised agents — a Crypto Agent processing market dynamics, a News Agent encoding weekly news sentiment, and a Trading Agent fusing all signals for portfolio execution — decompose the task across one of three communication architectures (hierarchical, collaborative, debate) and four capability configurations (zero-shot, \ac{cot}, \ac{rag}, skill-augmented). Each agent operates with a rolling memory window and a ReAct-style prompt that interleaves traceable reasoning with action generation. In a 52-week backtest over calendar year 2025 across the top 15 L1-blockchain native cryptocurrencies by market capitalisation as of January 2025, the best-performing configuration, \emph{Hierarchical (Skill)}, achieves a cumulative return of $+133.52\%$ and a Sharpe ratio of $+1.502$, substantially outperforming all single-agent variants, passive benchmarks, and deep learning baselines. A controlled ablation study identifies the Crypto Agent as the most critical component, with its removal reducing cumulative return by $42.57$ percentage points. A cross-model comparison further shows that \ac{mas} outperforms the single-agent baseline under GPT-4o, GPT-5, and Claude Sonnet 4.5, confirming that the benefit of multi-agent coordination is model-agnostic. Unlike black-box deep learning models, every portfolio decision in our framework is fully traceable to explicit agent reasoning chains, offering a principled, interpretable, and effective approach to multi-modal cryptocurrency portfolio management.

\end{abstract}


\section{Introduction}

Cryptocurrency portfolio management is an inherently multi-modal challenge: effective decision-making requires the simultaneous processing of structured market signals (price, volume, and on-chain time series), unstructured textual information (news and social media), and computed technical features~\cite{Jing2024AutomatedCandlesticks,Kapur2024CryptocurrencyLearning,Liu2021RisksCryptocurrency,LIU2022CommonCryptocurrency}, all under high volatility and limited asset pricing evidence~\cite{Corbet2019CryptocurrenciesAnalysis,Fang2022CryptocurrencySurvey}. Fusing these heterogeneous modalities while performing coherent, multi-step market reasoning~\cite{Hackethal2022TheInvestments} imposes a heavy cognitive workload, making expert cryptocurrency services either scarce or prohibitively expensive~\cite{CFPBoard2022CFPGuidelines}. Deep learning has been widely explored as an automation tool~\cite{Goutte2023DeepMarket,Lahmiri2019CryptocurrencyNetworks}, but its \enquote{black-box} nature raises concerns about trust and explainability that make investors hesitant to act on model outputs~\cite{Li2023PEN:Explainability,Carta2021ExplainableForecasting,Biran2017Human-centricPredictions}.

\Acp{llm} have emerged as a compelling alternative, offering the ability to process heterogeneous textual and structured data, generate traceable, human-readable rationales linking multi-modal observations to portfolio actions, and perform nuanced financial reasoning~\cite{Yang2024HarnessingBeyond,Li2023LargeSurvey,Wei2022Chain-of-ThoughtModels}. Nevertheless, single-\ac{llm} approaches remain limited in tasks requiring simultaneous multi-modal fusion and long-horizon reasoning~\cite{Xie2023PIXIU:Finance,Li2023MultimodalResearch}; this limitation is especially acute in cryptocurrency markets, where domain-specific knowledge is underrepresented in pretraining corpora. \Ac{mas} frameworks decompose complex tasks into modality-specialised subtasks handled by coordinated agents, offering a principled path forward~\cite{Wu2023AutoGen:Conversation,Pan2024AgentCoord:Collaboration}. Although \ac{mas} approaches have been explored in equity markets~\cite{Ding2024TradExpert:LLMs,Fatemi2024FinVision:Prediction,Oprea2025AAssets}, their application to cryptocurrency portfolio management remains limited to single-coin settings and single-modality data~\cite{Li2024ATrading,Yu2025FinNLP-FNP-LLMFinLegalChallenge}.

To fill this gap, we propose an \ac{llm}-powered \ac{mas} framework for automated cryptocurrency portfolio management. To support reproducibility and practical adoption, we open-source the full implementation at \url{https://anonymous.4open.science/r/cryptoMAS-FCB2/}. Our framework comprises three modality-specialised agents — a Crypto Agent processing structured market time series, a News Agent encoding unstructured weekly news sentiment, and a Trading Agent fusing all signals with portfolio state to produce trading actions — operating under one of three communication architectures (hierarchical, collaborative, debate) and one of four capability configurations (zero-shot, \ac{cot}, \ac{rag}, skill-augmented). All agents share a rolling memory window and a ReAct-style prompting structure~\cite{Yao2023ReAct:Models} that interleaves explicit reasoning with action generation, making every portfolio decision fully traceable and auditable. We conduct a 52-week backtest over calendar year 2025 across the top 15 L1-blockchain native cryptocurrencies by market capitalisation as of January 2025, benchmarking against passive hold strategies, classical technical indicators, five deep learning forecasters, and a single-agent \ac{llm} baseline. The main contributions of this paper are:

\begin{itemize}
    \item We propose a systematic multi-modal \ac{mas} framework for cryptocurrency portfolio management that jointly evaluates three communication architectures and four capability configurations, establishing the first comprehensive architecture-by-capability benchmark in this domain.

    \item We introduce three modality-specialised agents with rolling memory and ReAct-style prompting, and show that their hierarchical integration achieves a cumulative return of $+133.52\%$ and a Sharpe ratio of $+1.502$ over a 52-week backtest — outperforming all single-agent, deep learning, and passive baselines — with every decision traceable to explicit agent reasoning chains.

    \item We demonstrate that architecture and capability choices induce systematic trade-offs: skill augmentation maximises bull-market returns, \ac{cot} reasoning minimises bear-market drawdowns, and \ac{rag} grounding reduces volatility at the cost of upside capture.

    \item We conduct a controlled ablation study showing that the Crypto Agent is the primary driver of directional alpha, memory provides meaningful cross-week continuity, and the News Agent functions principally as a risk-dampening mechanism.

    \item We conduct a cross-model comparison among GPT-4o, GPT-5, and Claude Sonnet~4.5, demonstrating that \ac{mas} consistently outperforms the single-agent baseline under all three backbones and that Claude achieves the highest mean return across all 16 architecture--capability combinations, confirming the model-agnostic nature of the multi-agent coordination benefit.
\end{itemize}

\section{Related Work}

In this section, we examine works that use single \ac{llm} and multi-agent frameworks for investment. With their powerful text understanding and reasoning capabilities, \acp{llm} have become widely used in different investment tasks. Early studies have focused on employing single \ac{llm}s to predict asset prices and execute investment strategies. Some works have attempted to fine-tune their own financial \ac{llm} to complete investment tasks~\cite{Xie2023PIXIU:Finance,Liu2023FinGPT:Models,Li2023MultimodalResearch,Oprea2025AAssets}. Additionally, some studies specifically examined the performance of \ac{llm}s in trading three cryptocurrencies: Bitcoin, Ethereum, and Solana~\cite{Li2024ATrading,Yu2025FinNLP-FNP-LLMFinLegalChallenge}. However, the predictive power of single \ac{llm}s remains limited even after fine-tuning, and their results often exhibit significant bias. 

To further improve the performance of \ac{llm} in investment, recent research has shifted towards using multi-agent models for investment tasks. One notable example is the Summarize-Explain-Predict (SEP) framework, which employs a reflective agent that iteratively generates stock predictions and explanations with assistance from other agents~\cite{Koa2024LearningModels}. Some studies focus on using multiple agents to process data, summarize information, reflect, and generate stock predictions, respectively~\cite{Kou2024AutomateInvestment, Fatemi2024FinVision:Prediction, Ding2024TradExpert:LLMs}. However, there remains a gap in the development of multi-agent, multi-modal models specifically designed for cryptocurrency investment tasks. To fill this gap, we propose a multi-agent framework where specialized agents, each responsible for processing distinct modalities of information, collaboratively invest in a universe of leading cryptocurrencies.
\section{Methodology}

In this section, we first formalise the cryptocurrency portfolio management task and decompose it into a set of structured subtasks. We then introduce the three specialised agents at the core of our framework, describe the four capability configurations available to each agent, and present the multi-agent architectures that govern how agents interact and arrive at joint portfolio decisions. Finally, we outline the single-agent baseline and the portfolio execution model.

\subsection{Problem Formulation}
\label{subsec:problem_formulation}

We consider a universe $\mathcal{C} = \{c_i\}_{i=1}^{I}$ of $I = 15$ top L1-blockchain native cryptocurrencies by market capitalisation as of January 2025, fixed throughout the backtest to ensure liquidity.\footnote{BTC, ETH, BNB, XRP, SOL, TRX, ADA, BCH, HYPE, XMR, ZEC, LTC, SUI, AVAX, and HBAR.} At each ISO week $t$, the system observes two types of inputs: \begin{enumerate*} \item a set of market statistics $\mathbf{X}_{t-1} = \{\mathbf{x}_{c_i,t-1}\}_{i=1}^{I}$, where $\mathbf{x}_{c_i,t-1} \in \mathbb{R}^{n \times 3}$ contains the last $n = 30$ daily observations of closing price, trading volume, and market capitalisation for cryptocurrency $c_i$ sourced from CoinGecko, and \item a set of news articles $\mathbf{N}_{t-1} = \{N_{j,t-1}\}_{j=1}^{q}$, where $q$ denotes the number of relevant articles published during week $t-1$ sourced from Cointelegraph. \end{enumerate*}

Given these inputs, the system produces a vector of trading actions $\mathbf{A}_t = [a_{c_i,t}]_{i=1}^{I}$ for the current week, where each scalar action $a_{c_i,t} \in [-1, 1]$ specifies a fractional portfolio adjustment for asset $c_i$. Specifically, a positive value $a_{c_i,t} > 0$ instructs the system to allocate that fraction of the post-sell cash pool to purchase $c_i$, a negative value $a_{c_i,t} < 0$ instructs the system to liquidate that fraction of the current holding of $c_i$, and $a_{c_i,t} = 0$ indicates no change. All sell orders execute first; the resulting cash is then redistributed simultaneously across all assets with positive actions, with fractions scaled down proportionally if their sum exceeds unity. The overall objective is to maximise terminal portfolio value:
\begin{equation}
    \argmax_{\{\mathbf{A}_t\}_{t=1}^{T}} \; V_T, \qquad V_t = V_{t-1} \cdot \bigl(1 + r_t(\mathbf{A}_{t-1})\bigr),
    \label{eq:objective}
\end{equation}
where $r_t(\mathbf{A}_{t-1})$ denotes the portfolio return at week $t$ induced by actions $\mathbf{A}_{t-1}$, and the portfolio is initialised with a cash endowment of $V_0 = \$100{,}000$. A transaction cost of $0.1\%$ per trade side is applied to all executed orders to reflect realistic market frictions. Slippage and price impact are omitted: all 15 assets are top-ranked L1-blockchain native cryptocurrencies with combined daily spot trading volumes consistently exceeding billions of USD, ensuring that the weekly rebalancing orders executed by our strategy, sized relative to a \$100{,}000 portfolio, represent a negligible fraction of market depth and would not materially move prices.

\subsection{Framework Overview}
\label{subsec:framework_overview}

\begin{figure*}
    \centering
    \includegraphics[width=\linewidth]{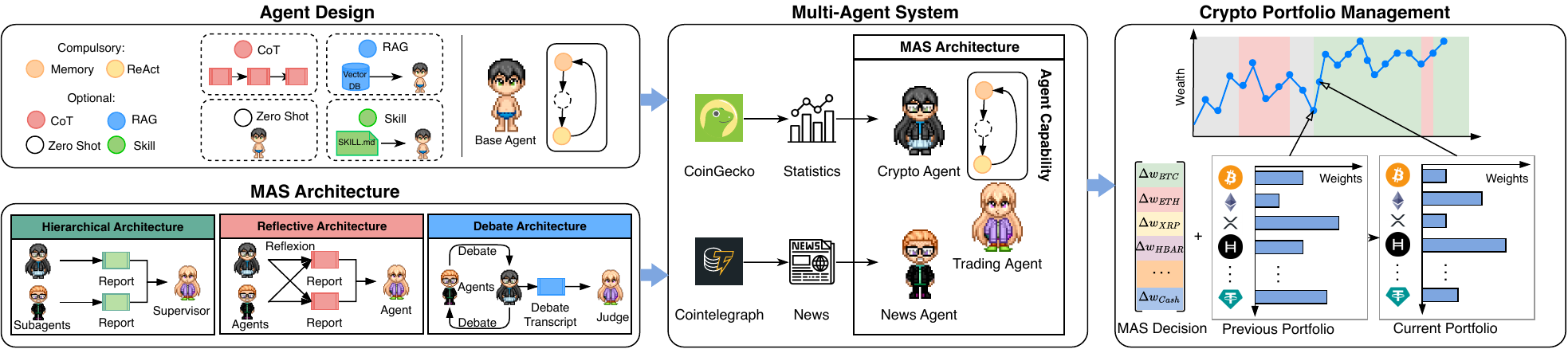}
    \caption{Multi-agent framework for automated cryptocurrency portfolio management.}
    \label{fig:flow}
\end{figure*}

We propose an \ac{llm}-powered multi-agent framework for cryptocurrency portfolio management, as illustrated in \autoref{fig:flow}. Our framework is organised along two orthogonal dimensions. The \emph{agent capability} dimension governs how individual agents process their inputs and reason about signals; four configurations are available: zero-shot, \ac{cot}, \ac{rag}, and skill-augmented. The \emph{\ac{mas} architecture} dimension determines how agents communicate and reconcile their views before the portfolio decision is issued; three architectures are evaluated: hierarchical, collaborative, and debate. All agents share two compulsory mechanisms across every configuration: a rolling memory window that grounds decisions in recent portfolio history, and a ReAct-style prompting structure~\cite{Yao2023ReAct:Models} that requires agents to interleave reasoning with action. Raw market statistics from CoinGecko and news articles from Cointelegraph are ingested at the start of each week, routed through the agent pipeline according to the selected architecture, and converted into trading actions that are applied at the weekly close.

\subsection{Agent Design}
\label{subsec:agent_design}

Our framework consists of three specialised agents that collectively decompose the portfolio management task into domain-specific subtasks, as illustrated in the central panel of \autoref{fig:flow}.

\subsubsection{Crypto Agent}
\label{subsubsec:crypto_agent}
The Crypto Agent is responsible for analysing recent market dynamics for each cryptocurrency and generating a directional signal for each asset. Given the 30-day market statistics $\mathbf{X}_{t-1}$ and the agent's rolling memory $\mathbf{M}^{\textnormal{mkt}}_{t-1}$, the Crypto Agent produces a signal array:
\begin{equation}
    \hat{\mathbf{S}}^{\textnormal{mkt}}_t = \textnormal{A}^{\textnormal{mkt}}\!\left(\mathbf{X}_{t-1},\, \mathbf{M}^{\textnormal{mkt}}_{t-1};\, \theta\right),
    \label{eq:crypto_agent}
\end{equation}
where $\hat{\mathbf{S}}^{\textnormal{mkt}}_t = [(s^{\textnormal{mkt}}_{c_i,t},\, \kappa^{\textnormal{mkt}}_{c_i,t},\, e^{\textnormal{mkt}}_{c_i,t})]_{i=1}^{I}$, with $s^{\textnormal{mkt}}_{c_i,t} \in [-1, 1]$ the directional signal ($-1 =$ strong bearish, $+1 =$ strong bullish), $\kappa^{\textnormal{mkt}}_{c_i,t} \in [0, 1]$ the associated confidence, and $e^{\textnormal{mkt}}_{c_i,t}$ a human-readable rationale.

\subsubsection{News Agent}
\label{subsubsec:news_agent}
The News Agent is responsible for processing the weekly Cointelegraph article set and producing an overall market sentiment signal together with coin-specific signals for individual cryptocurrencies explicitly mentioned in the news. Given the article set $\mathbf{N}_{t-1}$ and its rolling memory $\mathbf{M}^{\textnormal{news}}_{t-1}$, the News Agent produces:
\begin{equation}
    \hat{\mathbf{S}}^{\textnormal{news}}_t = \textnormal{A}^{\textnormal{news}}\!\left(\mathbf{N}_{t-1},\, \mathbf{M}^{\textnormal{news}}_{t-1};\, \theta\right),
    \label{eq:news_agent}
\end{equation}
where $\hat{\mathbf{S}}^{\textnormal{news}}_t = (s^{\textnormal{news}}_t,\, e^{\textnormal{news}}_t,\; [(s^{\textnormal{news}}_{c_i,t},\, \kappa^{\textnormal{news}}_{c_i,t},\, e^{\textnormal{news}}_{c_i,t})]_{c_i \in \mathcal{C}'})$, with $s^{\textnormal{news}}_t \in [-1, 1]$ the overall market sentiment, $e^{\textnormal{news}}_t$ an overall rationale, and $\mathcal{C}' \subseteq \mathcal{C}$ the subset of assets directly mentioned in the news. For assets outside $\mathcal{C}'$, the aggregate sentiment $s^{\textnormal{news}}_t$ is used as a fallback signal.

\subsubsection{Trading Agent}
\label{subsubsec:trading_agent}
The Trading Agent integrates the outputs of the Crypto Agent and the News Agent together with the current portfolio state $\Pi_{t-1}$, encoding the cash balance, per-asset holdings, and unrealised profit and loss for each position, to produce the final trading actions:
\begin{equation}
    \mathbf{A}_t = \textnormal{A}^{\textnormal{trade}}\!\left(\hat{\mathbf{S}}^{\textnormal{mkt}}_t,\, \hat{\mathbf{S}}^{\textnormal{news}}_t,\, \Pi_{t-1},\, \mathbf{M}^{\textnormal{trade}}_{t-1};\, \theta\right).
    \label{eq:trading_agent}
\end{equation}
The Trading Agent is instructed to avoid excessive concentration in any single asset, to use per-position unrealised returns to inform profit-taking and loss-cutting decisions, and to act conservatively when signals are conflicting or confidence is low.

\subsection{Agent Capabilities}
\label{subsec:capabilities}

Each agent can be equipped with one of four capability configurations that augment how it processes its inputs and reasons about signals, as illustrated in the left panel of \autoref{fig:flow}. Two mechanisms are compulsory across all configurations: a rolling memory window (detailed in \autoref{subsec:memory}) and a ReAct-style prompt structure. The four optional capability configurations are as follows.

\subsubsection{Zero-Shot}
\label{subsubsec:zero_shot}
In the zero-shot configuration, no additional reasoning scaffolding is provided beyond the compulsory memory and ReAct mechanisms. Each agent maps its inputs directly to structured outputs in a single forward pass, relying entirely on the in-context role description and the pretrained knowledge of the underlying \ac{llm}. This configuration serves as the capability-level baseline.

\subsubsection{Chain-of-Thought}
\label{subsubsec:cot}
The \ac{cot} configuration~\cite{Wei2022Chain-of-ThoughtModels} augments each agent's system prompt with an explicit instruction to reason step-by-step before producing its final structured output. The agent is required to emit its intermediate reasoning inside the tags \texttt{<reasoning>...</reasoning>}; the structured \ac{json} output must appear after the closing tag. This design forces the model to externalise its inference chain, which has been shown to improve performance on complex multi-step financial reasoning tasks~\cite{Koa2024LearningModels}. When the model erroneously places the \ac{json} inside the reasoning block, the parser falls back to recovering it from within the tags.

\subsubsection{Retrieval-Augmented Generation}
\label{subsubsec:rag}
The \ac{rag} configuration~\cite{Lewis2020Retrieval-AugmentedTasks} augments the Crypto Agent with a pre-built vector store of historical market analogues. For each asset $c_i$, the current market snapshot is encoded into an 11-dimensional scale-invariant feature vector:
\begin{equation}
    \small
    \mathbf{f}_{c_i,t} = \bigl[\underbrace{r^{\textnormal{px}}_{1d},\; r^{\textnormal{px}}_{3d},\; r^{\textnormal{px}}_{7d},\; r^{\textnormal{px}}_{14d},\; r^{\textnormal{px}}_{21d},\; r^{\textnormal{px}}_{28d}}_{\text{price returns}},\; \underbrace{r^{\textnormal{vol}}_{7d},\; r^{\textnormal{vol}}_{14d},\; r^{\textnormal{vol}}_{28d}}_{\text{volume changes}},\; \underbrace{r^{\textnormal{mcap}}_{7d},\; r^{\textnormal{mcap}}_{28d}}_{\text{mcap changes}}\bigr]^{\!\top},
    \label{eq:rag_features}
\end{equation}
where $r^{\textnormal{px}}_{kd}$, $r^{\textnormal{vol}}_{kd}$, and $r^{\textnormal{mcap}}_{kd}$ denote the $k$-day percentage changes in price, trading volume, and market capitalisation, respectively. All feature vectors are L2-normalised; retrieval is performed via cosine similarity against the historical store, returning the top-$K = 3$ most similar past weeks for each asset along with their realised next-week returns. To prevent lookahead bias, only historical entries strictly before the current week $t$ are eligible for retrieval. The retrieved analogues are injected into the Crypto Agent's prompt as empirical calibration points, supplementing the raw market data with outcome-labelled historical precedents.

\subsubsection{Skill-Augmented}
\label{subsubsec:skill}
The skill-augmented configuration provides the Crypto Agent with pre-computed technical indicator signals derived from four classical trading strategies. For each asset $c_i$, the following indicators are computed from the 30-day price history $\mathbf{x}_{c_i,t-1}$ and injected into the prompt alongside the raw market data:

\begin{itemize}
    \item \textbf{SMA\textsubscript{7}}: A bullish signal is generated when the current closing price $p_t$ exceeds the 7-day simple moving average $\bar{p}_{7}$, i.e., $p_t > \bar{p}_{7}$, indicating that price is trending above its short-term average.
    \item \textbf{SLMA}: A bullish signal is generated when $\bar{p}_{7} > \bar{p}_{30}$, i.e., the short-term moving average has crossed above the long-term moving average, a classical golden-cross condition.
    \item \textbf{MACD}: A bullish signal is generated when the MACD histogram $h_t = (\textnormal{EMA}_{12} - \textnormal{EMA}_{26}) - \textnormal{EMA}_{9}(\textnormal{EMA}_{12} - \textnormal{EMA}_{26}) > 0$, indicating upward price momentum.
    \item \textbf{BB}: A contrarian bullish signal is generated when $p_t < \bar{p}_{20} - 2\sigma_{20}$, where $\sigma_{20}$ is the 20-day rolling standard deviation, indicating an oversold condition relative to the recent price distribution and a potential mean-reversion opportunity.
\end{itemize}

For each asset, the four binary readings are summarised in a composite signal table alongside their numerical values and a bullish count $b_{c_i,t} = \sum_{k} \mathbf{1}[\text{indicator}_k \text{ is bullish}] \in \{0, 1, 2, 3, 4\}$. The Crypto Agent's system prompt explicitly maps $b_{c_i,t}$ to signal strength: $b_{c_i,t} = 4$ corresponds to a strong buy, $b_{c_i,t} = 0$ to a strong sell, and intermediate values to proportionally moderate stances. This configuration bridges the gap between interpretable rule-based strategies and \ac{llm}-based reasoning by providing market priors that the agent can incorporate alongside its broader contextual understanding.

\subsection{Agent Memory}
\label{subsec:memory}

All agents maintain a rolling memory of their $K = 4$ most recent weekly outputs. Before each new call, the memory block is prepended to the user message in reverse chronological order, enabling the agent to reason about trends across weeks, maintain signal consistency, and detect patterns such as momentum continuation or regime change. Each memory entry records the ISO week identifier and the agent's complete structured output for that week. Memory is serialised at the end of each week and restored at the start of the next, supporting full experiment resumption without loss of context. When operating in refinement or debate rounds within the collaborative and debate architectures, the agent's memory is updated only with the final-round output so that intermediate reasoning steps do not pollute the longitudinal record.

\subsection{Multi-Agent Architectures}
\label{subsec:architectures}

We evaluate three \ac{mas} architectures that differ in how the Crypto Agent and the News Agent communicate before the Trading Agent issues its final decision, as illustrated in the right panel of \autoref{fig:flow}. Let $\mathbf{S}^{\textnormal{mkt},(r)}_t$ and $\mathbf{S}^{\textnormal{news},(r)}_t$ denote the outputs of the Crypto Agent and the News Agent after communication round $r$.

\subsubsection{Hierarchical Architecture}
\label{subsubsec:hierarchical}
In the hierarchical architecture, the Crypto Agent and the News Agent operate as independent subordinate analysts, each completing a single-pass analysis of its domain-specific inputs~\cite{Hong2023MetaGPT:Framework}. Their respective signal reports $\mathbf{S}^{\textnormal{mkt}}_t$ and $\mathbf{S}^{\textnormal{news}}_t$ are forwarded to the Trading Agent, which acts as a supervisory agent. The Trading Agent's system prompt explicitly frames it as a supervisor integrating structured reports from specialised subordinates, and it bears sole responsibility for reconciling conflicting signals and issuing the final trading actions $\mathbf{A}_t$. This architecture mirrors the command-and-control structure common in institutional trading desks, where specialist analysts report to a portfolio manager who has ultimate decision authority.

\subsubsection{Collaborative Architecture}
\label{subsubsec:collaborative}
In the collaborative architecture, the analytical agents engage in iterative cross-agent refinement before the Trading Agent produces its decision. In round $r = 0$, both agents produce independent initial signals without inter-agent communication. In each subsequent refinement round $r \geq 1$, the Crypto Agent receives the News Agent's most recent output as contextual information and is invited to revise its signals accordingly; the News Agent then receives the Crypto Agent's revised output and similarly reflects. Formally:
\begin{align}
    \mathbf{S}^{\textnormal{mkt},(r)}_t &= \textnormal{A}^{\textnormal{mkt}}\!\left(\mathbf{X}_{t-1},\; \mathbf{S}^{\textnormal{news},(r-1)}_t\right), \label{eq:collab_mkt} \\
    \mathbf{S}^{\textnormal{news},(r)}_t &= \textnormal{A}^{\textnormal{news}}\!\left(\mathbf{N}_{t-1},\; \mathbf{S}^{\textnormal{mkt},(r)}_t\right). \label{eq:collab_news}
\end{align}
After $R = 1$ refinement round, the converged signals are committed to each agent's memory, and the Trading Agent integrates them to produce $\mathbf{A}_t$. This architecture draws on the reflexion paradigm~\cite{Shinn2023Reflexion:Learning}, where agents self-improve by incorporating cross-modal signals into their analysis without requiring an explicit supervisor.

\subsubsection{Debate Architecture}
\label{subsubsec:debate}
In the debate architecture, the analytical agents engage in structured adversarial argumentation before the Trading Agent acts as a judge. In round $r = 0$, both agents independently form initial positions. In each subsequent debate round $r \geq 1$, each agent is explicitly instructed to challenge the opponent's most recent argument, cite specific evidence from the data in support of its position, and produce an updated signal that justifies any revision or maintained disagreement:
\begin{equation}
    \mathbf{S}^{(r)}_t = \textnormal{A}\!\left(\mathbf{D}_{t-1},\; \mathbf{S}^{(r-1)}_{\textnormal{opp},t},\; \mathbf{S}^{(r-1)}_{\textnormal{own},t}\right),
    \label{eq:debate}
\end{equation}
where $\mathbf{D}_{t-1}$ denotes the agent's primary data inputs and $\mathbf{S}^{(r-1)}_{\textnormal{opp},t}$ denotes the opponent's most recent argument. After $R = 2$ debate rounds, the full transcript, comprising all rounds of both agents' evolving positions, is forwarded to the Trading Agent. The Trading Agent is instructed to weigh the strength of each side's arguments, identify points of convergence and persistent disagreement, and produce $\mathbf{A}_t$ based on its evaluation of the complete dialectical record. This architecture is motivated by the observation that adversarial deliberation can surface hidden assumptions and contradictions in agent reasoning that cooperative exchanges may leave unexamined~\cite{Du2023ImprovingDebate}.

\subsection{Single Agent versus Multi-Agent System}
\label{subsec:single_vs_mas}

To establish a rigorous baseline, we compare our \ac{mas} against a single-agent model. The single-agent model receives all available inputs, market statistics $\mathbf{X}_{t-1}$, news articles $\mathbf{N}_{t-1}$, and the current portfolio state $\Pi_{t-1}$, in a single \ac{llm} call and directly produces the trading actions $\mathbf{A}_t$, bypassing the specialised agent pipeline entirely. In contrast, our \ac{mas} decomposes this task across three agents: the Crypto Agent and the News Agent process domain-specific inputs independently, and the Trading Agent integrates their outputs with the portfolio context to form the final decision. As noted in~\cite{Koa2024LearningModels}, single \acp{llm} often struggle to effectively weight and integrate heterogeneous information modalities when forming aggregate predictions in financial contexts. Our framework addresses this by assigning each modality to a dedicated agent with a domain-specific prompt and enabling structured inter-agent communication — hierarchical reporting, collaborative refinement, or adversarial debate — that surfaces informational conflicts before the final decision.

\section{Experiment}

In this section, we evaluate the portfolio performance of our \ac{mas} framework against a comprehensive set of baselines across the full backtest period and across distinct market regimes, and conduct an ablation study to isolate the contribution of each system component.

\subsection{Experiment Settings}
\label{subsec:experiment_settings}

\subsubsection{Data and Backtest Setup}
We employ \textbf{GPT-4o}, \textbf{GPT-5}, and \textbf{Claude Sonnet~4.5} as backbone \ac{llm}s for all agent-powered strategies, enabling a cross-model assessment of how backbone choice affects portfolio performance across architectures and capability configurations.\footnote{\url{https://platform.openai.com/docs/models}; \url{https://www.anthropic.com/claude}. All experiments run on a dual AMD EPYC 7F32 server (128\,GB DDR4 ECC RAM). \ac{llm} inference is via the OpenAI API (GPT-4o, GPT-5) and the Anthropic API (Claude Sonnet~4.5) at temperature $0.0$; no local GPU resources are required.} Market data is sourced from CoinGecko and news articles from Cointelegraph. The backtest spans the full calendar year 2025, from the first to the last ISO week ($T = 52$ weeks), over a universe $\mathcal{C}$ of the top 15 L1-blockchain native cryptocurrencies by market capitalisation as of January 2025.\footnote{BTC, ETH, BNB, XRP, SOL, TRX, ADA, BCH, HYPE, XMR, ZEC, LTC, SUI, AVAX, and HBAR. The universe is fixed at the start of the backtest to ensure liquidity.} This window is selected to ensure a leakage-free evaluation: GPT-4o has a training-data cutoff of October 2023, GPT-5 of September 2024, and Claude Sonnet~4.5 of January 2025, so the evaluation period lies strictly outside the pre-training distribution of all three models.

Each strategy is initialised with $V_0 = \$100{,}000$; a transaction cost of $0.1\%$ per trade side is applied to all executed orders; slippage and price impact are abstracted away given the exceptional liquidity of the selected assets. Market regimes are identified using the methodology of~\cite{Cagan2024StockMarket}, which classifies each week based on the mcap-weighted basket return relative to its running peak and trough: weeks in which the basket trades at or above $+20\%$ of its running trough are classified as \emph{bull} ($N_{\textnormal{bull}} = 27$ weeks); weeks in which it trades at or below $-20\%$ of its running peak are classified as \emph{bear} ($N_{\textnormal{bear}} = 15$ weeks).

\subsubsection{Baselines}
\label{subsubsec:baselines}
We compare our framework against three groups of baselines:

\begin{itemize}
    \item \textbf{Hold}: We construct a 100\% Bitcoin buy-and-hold portfolio (\emph{BTC Hold}) and a market-capitalisation-weighted buy-and-hold portfolio across all 15 assets (\emph{MCap Hold}).


    \item \textbf{Deep Learning}: We implement five neural time-series forecasters, \emph{LSTM}~\cite{Hochreiter1997LongMemory}, \emph{Informer}~\cite{Zhou2021Informer:Forecasting}, \emph{Autoformer}~\cite{Wu2021Autoformer:Forecasting}, \emph{TimesNet}~\cite{Wu2023TimesNet:Analysis}, and \emph{PatchTST}~\cite{Nie2023ATransformers}, each trained on a daily multivariate feature set, comprising close price, trading volume, and market capitalisation, for all assets in $\mathcal{C}$ prior to 2025, matching the quantitative market signals available to the Crypto Agent. To align the prediction horizon with the weekly rebalancing period, all models are configured to forecast the 7-day-ahead close price: LSTM is trained with a target of $\text{close}_{t+7}$ given the preceding 5 days of multivariate inputs, while the transformer-based models use a forecast horizon of $h=7$ with a 32-day multivariate lookback. A long signal is generated for each asset whose predicted 7-day-ahead close exceeds its current close; positions are equally weighted across all signalled assets and fully rebalanced each week.

    \item \textbf{Agent}: We include a single-agent \ac{llm} baseline (\emph{SA}) that receives raw market data, news articles, and the current portfolio state in a single call and directly produces trading actions, under all four capability configurations described in \autoref{subsec:capabilities}. This baseline isolates the contribution of the multi-agent architecture from the choice of prompting strategy.
\end{itemize}

\subsubsection{Evaluation Metrics}
\label{subsubsec:metrics}
Following standard practice in empirical asset pricing~\cite{Asness2013ValueEverywhere, Gu2020EmpiricalLearning, Sharpe1994TheRatio}, we evaluate each strategy on six metrics. Let $r_t$ denote the portfolio's weekly return at week $t$, $T$ the total number of weeks in the evaluation window, $r_f$ the weekly risk-free rate (one-month US Treasury bill from the Fama-French data library\footnote{\url{https://mba.tuck.dartmouth.edu/pages/faculty/ken.french/data_library.html}}), and $V_t$ the portfolio value at week $t$.

\begin{itemize}
    \item \textbf{Cumulative Return (Cum\%)}~\cite{Asness2013ValueEverywhere} measures total portfolio growth over the evaluation period: $\prod_{t=1}^{T}(1 + r_t) - 1$.
    \item \textbf{Average Weekly Return (Avg\%)}~\cite{Gu2020EmpiricalLearning} measures the mean of $r_t$ over the evaluation window.
    \item \textbf{Annualised Volatility (Vol\%)}~\cite{Gu2020EmpiricalLearning} measures the standard deviation of weekly returns scaled to one year: $\sigma_w \cdot \sqrt{52}$, where $\sigma_w = \sqrt{\frac{1}{T-1}\sum_{t=1}^{T}(r_t - \bar{r})^2}$.
    \item \textbf{Sharpe Ratio (SR)}~\cite{Sharpe1994TheRatio} measures the annualised risk-adjusted excess return: $(\bar{r} - r_f) \cdot \sqrt{52} / \sigma_w$.
    \item \textbf{Maximum Drawdown (MDD\%)}~\cite{Gu2020EmpiricalLearning} measures the largest peak-to-trough decline in portfolio value: $\min_{t}\!\left(V_t / \max_{s \leq t} V_s - 1\right)$.
    \item \textbf{Win Rate (Win\%)}~\cite{Luo2026ResistingReasoning} is the fraction of weeks with a positive return: $\frac{1}{T}\sum_{t=1}^{T}\mathbf{1}[r_t > 0]$.
\end{itemize}

\subsection{Performance Comparison}
\label{subsec:performance}

\begin{figure*}
    \centering
    \includegraphics[width=\linewidth]{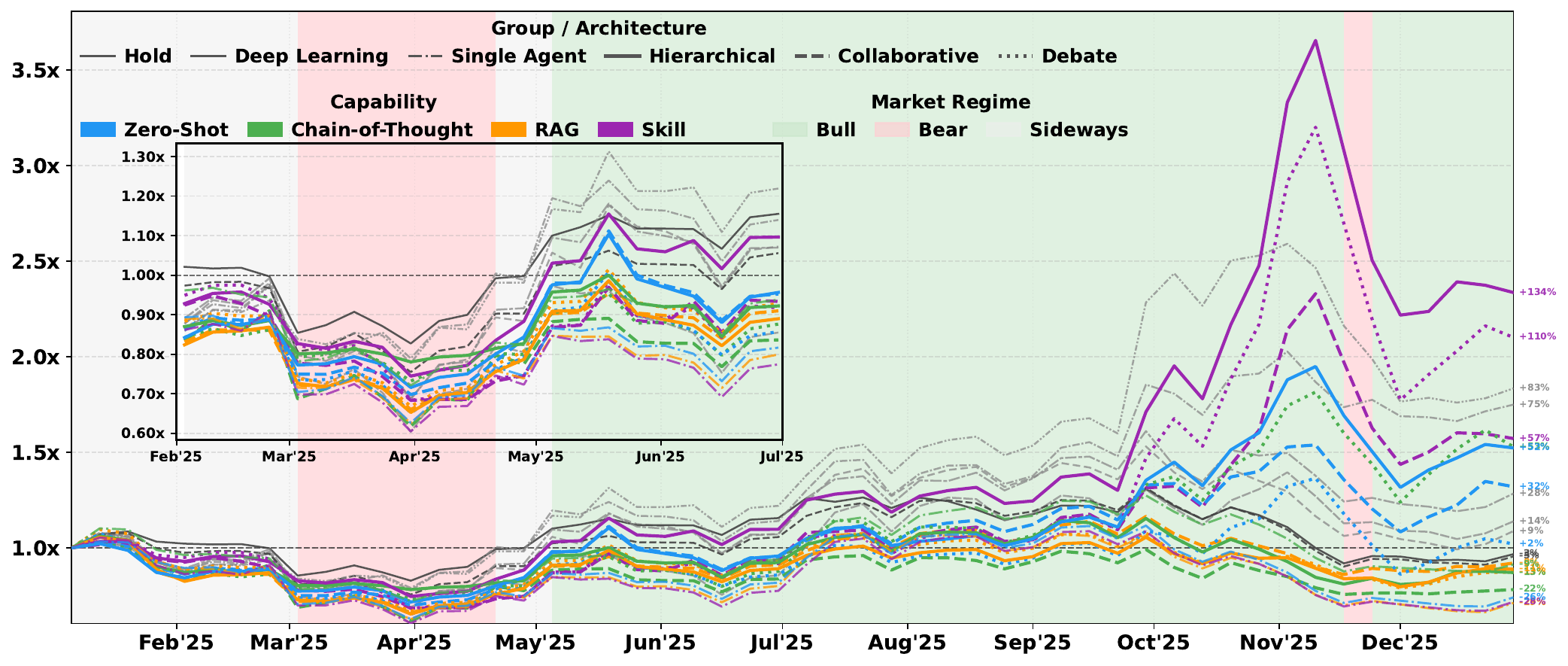}
    \caption{Performance comparisons in out-of-sample cumulative returns of our multi-agent model portfolio against baselines (GPT-4o backbone).}
    \label{fig:portfolio}
\end{figure*}
\begin{table*}[ht]
\centering
\caption{Strategy performance across market regimes (GPT-4o backbone). Each cell is colored to indicate the \colorbox{orange!60}{best}, \colorbox{orange!30}{second best}, and \colorbox{yellow!30}{third best} performing strategy per metric and regime.}
\label{tab:performance}
\scriptsize
\setlength{\tabcolsep}{4pt}
\resizebox{\linewidth}{!}{
\begin{tabular}{c l | rrrrrr | rrrrrr | rrrrrr}
\toprule
\multicolumn{2}{c|}{\multirow{2}{*}{\textbf{Strategy}}} & \multicolumn{6}{c|}{\textbf{All} (full period, $N=52$)} & \multicolumn{6}{c|}{\textbf{Bull} (rally $\geq+20\%$, $N=27$)} & \multicolumn{6}{c}{\textbf{Bear} (drawdown $\leq-20\%$, $N=15$)} \\
\cmidrule(lr){3-8} \cmidrule(lr){9-14} \cmidrule(l){15-20}
& & Cum\%\textcolor{red}{$\uparrow$} & Avg\%\textcolor{red}{$\uparrow$} & Vol\%\textcolor{teal}{$\downarrow$} & SR\textcolor{red}{$\uparrow$} & MDD\%\textcolor{red}{$\uparrow$} & Win\%\textcolor{red}{$\uparrow$} & Cum\%\textcolor{red}{$\uparrow$} & Avg\%\textcolor{red}{$\uparrow$} & Vol\%\textcolor{teal}{$\downarrow$} & SR\textcolor{red}{$\uparrow$} & MDD\%\textcolor{red}{$\uparrow$} & Win\%\textcolor{red}{$\uparrow$} & Cum\%\textcolor{red}{$\uparrow$} & Avg\%\textcolor{red}{$\uparrow$} & Vol\%\textcolor{teal}{$\downarrow$} & SR\textcolor{red}{$\uparrow$} & MDD\%\textcolor{red}{$\uparrow$} & Win\%\textcolor{red}{$\uparrow$} \\
\midrule
\multicolumn{1}{c}{\multirow{2}{*}{\rotatebox[origin=c]{90}{{\tiny\textbf{Hold}}}}} & BTC Hold & -3.36 & +0.06 & \cellcolor{orange!60}37.09 & +0.091 & \cellcolor{orange!30}-29.73 & 46.2 & +11.01 & +0.50 & \cellcolor{orange!60}35.63 & +0.734 & -15.22 & 44.4 & \cellcolor{orange!30}-21.24 & \cellcolor{orange!30}-1.41 & \cellcolor{yellow!30}42.62 & -1.719 & \cellcolor{orange!30}-8.83 & 46.7 \\
 & MCap Hold & -4.54 & +0.08 & \cellcolor{yellow!30}41.65 & +0.095 & -31.31 & 50.0 & +21.13 & +0.86 & \cellcolor{orange!30}40.82 & +1.100 & -16.67 & 51.9 & -26.00 & -1.78 & 46.82 & -1.976 & -11.52 & 40.0 \\
\cline{1-20}
\multicolumn{1}{c}{\multirow{5}{*}{\rotatebox[origin=c]{90}{{\tiny\textbf{Deep Learning}}}}} & LSTM & +13.82 & +0.67 & 66.37 & +0.522 & -44.01 & 48.1 & +73.16 & +2.45 & 66.47 & +1.916 & -22.59 & 55.6 & -40.64 & -2.97 & 68.75 & -2.243 & -20.80 & 40.0 \\
 & Informer & +74.89 & +1.40 & 58.77 & \cellcolor{yellow!30}+1.236 & \cellcolor{orange!60}-26.37 & 53.8 & +105.12 & +3.06 & 64.40 & +2.470 & -16.51 & 63.0 & -22.24 & -1.45 & 47.61 & \cellcolor{orange!30}-1.583 & -10.95 & 46.7 \\
 & Autoformer & +8.72 & +0.48 & 58.31 & +0.428 & -35.22 & 46.2 & +46.45 & +1.75 & 60.83 & +1.499 & -17.96 & 51.9 & -32.26 & -2.24 & 58.05 & -2.011 & -14.88 & 40.0 \\
 & TimesNet & \cellcolor{yellow!30}+83.40 & \cellcolor{yellow!30}+1.55 & 66.72 & +1.212 & -32.09 & 55.8 & +164.04 & +4.14 & 76.93 & +2.799 & -15.53 & \cellcolor{yellow!30}70.4 & -34.73 & -2.60 & 47.24 & -2.862 & -28.63 & 33.3 \\
 & PatchTST & +28.35 & +0.81 & 59.43 & +0.710 & -35.68 & 50.0 & +63.20 & +2.16 & 60.92 & +1.846 & -17.75 & 55.6 & -32.47 & -2.28 & 56.30 & -2.110 & -15.71 & 40.0 \\
\cline{1-20}
\multicolumn{1}{c}{\multirow{4}{*}{\rotatebox[origin=c]{90}{{\tiny\textbf{Agent}}}}} & SA (ZS) & -26.00 & -0.31 & 52.68 & -0.309 & -39.29 & 50.0 & +16.69 & +0.84 & 54.77 & +0.802 & -21.99 & 59.3 & -33.53 & -2.37 & 57.10 & -2.162 & -15.62 & 40.0 \\
 & SA (CoT) & -8.81 & +0.01 & 44.54 & +0.013 & -34.48 & 46.2 & +27.20 & +1.08 & 45.77 & +1.232 & -17.76 & 55.6 & -28.85 & -2.05 & 44.58 & -2.395 & -11.80 & 40.0 \\
 & SA (RAG) & -28.90 & -0.39 & 52.69 & -0.386 & -40.62 & 44.2 & +15.09 & +0.79 & 54.73 & +0.751 & -21.02 & 51.9 & -35.91 & -2.64 & 54.58 & -2.513 & -14.86 & 33.3 \\
 & SA (Skill) & -28.34 & -0.33 & 56.81 & -0.302 & -42.55 & 48.1 & +17.35 & +0.89 & 57.69 & +0.805 & -21.87 & 55.6 & -37.90 & -2.77 & 60.03 & -2.404 & -16.53 & 40.0 \\
\cdashline{1-20}
\multicolumn{1}{c}{\multirow{12}{*}{\rotatebox[origin=c]{90}{{\tiny\textbf{Multi-Agent System}}}}} & Hier.\ (ZS) & +52.19 & +1.10 & 56.24 & +1.022 & -32.47 & \cellcolor{orange!60}61.5 & +122.69 & +3.33 & 59.99 & +2.885 & -20.24 & 70.4 & -31.54 & -2.24 & 52.28 & -2.225 & -32.47 & 53.3 \\
 & Hier.\ (CoT) & -12.99 & -0.10 & 42.24 & -0.122 & \cellcolor{yellow!30}-29.98 & 53.8 & +12.09 & +0.64 & 49.28 & +0.680 & -19.59 & 55.6 & \cellcolor{orange!60}-15.51 & \cellcolor{orange!60}-1.03 & \cellcolor{orange!60}31.25 & -1.709 & \cellcolor{orange!60}-4.42 & \cellcolor{yellow!30}53.3 \\
 & Hier.\ (RAG) & -11.25 & -0.05 & 43.63 & -0.056 & -38.11 & 57.7 & +20.37 & +0.86 & 42.80 & +1.039 & -16.67 & 63.0 & -24.14 & -1.64 & 43.71 & -1.956 & -11.39 & 46.7 \\
 & Hier.\ (Skill) & \cellcolor{orange!60}+133.52 & \cellcolor{orange!60}+2.12 & 73.40 & \cellcolor{orange!60}+1.502 & -39.33 & 59.6 & \cellcolor{orange!30}+277.08 & \cellcolor{orange!30}+5.60 & 82.27 & \cellcolor{orange!60}+3.541 & \cellcolor{orange!30}-12.00 & \cellcolor{orange!60}74.1 & -41.54 & -3.19 & 58.86 & -2.814 & -39.33 & 40.0 \\
 & Collab.\ (ZS) & +31.91 & +0.80 & 52.84 & +0.785 & -32.19 & \cellcolor{orange!30}59.6 & +82.07 & +2.51 & 54.02 & +2.413 & -20.35 & 70.4 & -29.27 & -2.00 & 54.79 & -1.900 & -29.60 & 46.7 \\
 & Collab.\ (CoT) & -21.94 & -0.22 & 51.48 & -0.220 & -39.73 & 48.1 & +9.46 & +0.52 & 45.46 & +0.598 & -15.31 & 48.1 & -30.84 & -2.09 & 58.61 & -1.851 & -17.18 & \cellcolor{orange!60}53.3 \\
 & Collab.\ (RAG) & -8.03 & -0.00 & \cellcolor{orange!30}40.44 & -0.006 & -35.22 & 51.9 & +22.20 & +0.90 & \cellcolor{yellow!30}41.27 & +1.134 & -16.73 & 59.3 & \cellcolor{yellow!30}-22.13 & -1.50 & \cellcolor{orange!30}40.08 & -1.946 & \cellcolor{yellow!30}-10.42 & 40.0 \\
 & Collab.\ (Skill) & +57.12 & +1.28 & 67.41 & +0.990 & -38.35 & 57.7 & \cellcolor{yellow!30}+187.14 & \cellcolor{yellow!30}+4.46 & 74.49 & +3.112 & \cellcolor{orange!60}-11.75 & 70.4 & -43.87 & -3.45 & 58.32 & -3.080 & -38.35 & 40.0 \\
 & Debate (ZS) & +1.94 & +0.35 & 57.47 & +0.320 & -39.20 & 57.7 & +66.85 & +2.18 & 54.75 & +2.075 & -20.79 & 63.0 & -39.51 & -2.89 & 65.21 & -2.306 & -36.84 & \cellcolor{orange!30}53.3 \\
 & Debate (CoT) & +52.76 & +1.11 & 56.37 & +1.026 & -31.72 & 53.8 & +117.97 & +3.24 & 59.71 & +2.826 & -16.34 & 63.0 & -22.84 & \cellcolor{yellow!30}-1.42 & 56.46 & \cellcolor{orange!60}-1.311 & -31.72 & 46.7 \\
 & Debate (RAG) & -12.71 & -0.07 & 44.35 & -0.085 & -38.97 & 55.8 & +18.21 & +0.79 & 43.22 & +0.953 & -15.82 & 63.0 & -27.16 & -1.90 & 44.51 & -2.220 & -13.62 & 46.7 \\
 & Debate (Skill) & \cellcolor{orange!30}+110.00 & \cellcolor{orange!30}+2.00 & 79.26 & \cellcolor{orange!30}+1.310 & -44.60 & \cellcolor{yellow!30}59.6 & \cellcolor{orange!60}+290.96 & \cellcolor{orange!60}+5.81 & 86.89 & \cellcolor{orange!30}+3.475 & -14.51 & 70.4 & -47.30 & -3.71 & 70.27 & -2.745 & -44.60 & 40.0 \\
\bottomrule
\end{tabular}
}
\end{table*}
\begin{figure}
    \centering
    \includegraphics[width=\linewidth]{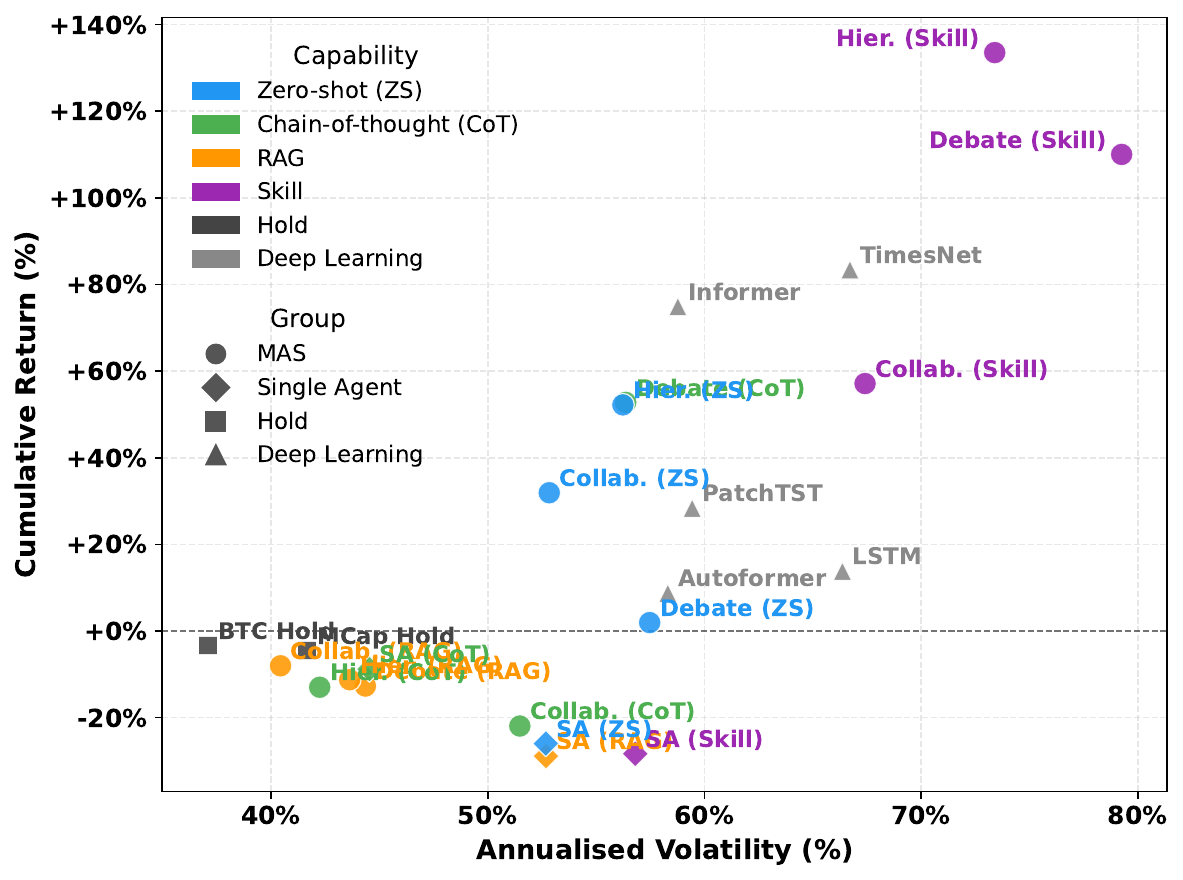}
    \caption{Risk--return scatter plot for all strategies over the full 52-week backtest (GPT-4o backbone).}
    \label{fig:risk_return}
\end{figure}

\autoref{tab:performance} reports strategy performance across the full period and the bull and bear regimes; all \ac{llm}-powered strategies in this table use GPT-4o as the backbone. We make the following observations.

\subsubsection{Overall Portfolio Performance}
\label{subsubsec:overall}

\Ac{mas} strategies substantially outperform both the hold benchmarks and the single-agent baseline across the full evaluation period. The top-performing strategy is \emph{Hierarchical (Skill)}, which achieves a cumulative return of $+133.52\%$ and a Sharpe ratio of $+1.502$, surpassing the best non-\ac{mas} strategy (\emph{TimesNet}, $+83.40\%$, SR $= +1.212$) by more than 50 percentage points in cumulative return. The second-ranked strategy, \emph{Debate (Skill)}, delivers $+110.00\%$ at a Sharpe ratio of $+1.310$. In contrast, all four single-agent variants record negative or near-zero cumulative returns, with the best \ac{sa} variant (\emph{SA (CoT)}) returning $-8.81\%$. The hold benchmarks are similarly negative ($-3.36\%$ and $-4.54\%$ for BTC Hold and MCap Hold, respectively), confirming that 2025 presented a challenging environment in which passive strategies underperformed. These results indicate that the multi-agent decomposition, assigning domain-specific reasoning to specialised agents and enabling structured inter-agent communication, provides a decisive advantage over both single-agent and passive approaches.

Among capability variants, skill augmentation consistently yields the highest returns across all three \ac{mas} architectures in the full period, with Hierarchical (Skill), Debate (Skill), and Collaborative (Skill) ranking first, second, and fifth overall, respectively. The \ac{rag} configuration achieves the lowest annualised volatility across all strategies ($40.44\%$ for Collaborative (RAG)), indicating that grounding signals in historical analogues promotes more conservative and consistent decision-making. The zero-shot and \ac{cot} configurations show mixed results: Hierarchical (ZS) achieves a respectable $+52.19\%$ and Debate (CoT) reaches $+52.76\%$, while their collaborative counterparts perform more modestly.

\autoref{fig:risk_return} visualises the risk--return trade-off across all strategies. \Ac{mas} configurations (circles) form a distinct upper cluster clearly separated from single-agent (diamonds), deep learning (triangles), and hold (squares) groups, which all lie near or below the zero-return line. Within the \ac{mas} cluster, skill-augmented strategies carry the highest volatility but also the highest returns, while \ac{rag} variants shift leftward, reflecting their conservatism.

\subsubsection{Regime-Conditioned Analysis}
\label{subsubsec:regimes}

The regime breakdown reveals that strategy rankings vary considerably across market conditions, motivating the importance of evaluating beyond the full-period aggregate.

\paragraph{Bull Market.} During the 27-week bull period, skill-augmented \ac{mas} strategies dominate. \emph{Debate (Skill)} delivers the highest cumulative return of $+290.96\%$ (SR $= +3.475$), followed by \emph{Hierarchical (Skill)} at $+277.08\%$ (SR $= +3.541$) and \emph{Collaborative (Skill)} at $+187.14\%$ (SR $= +3.112$). Notably, \emph{Hierarchical (Skill)} leads on risk-adjusted terms despite ranking second by cumulative return, reflecting its lower realised volatility ($82.27\%$ versus $86.89\%$ for Debate (Skill)). The pre-computed technical indicators provide agents with directional priors that closely align with trending market conditions, driving aggressive and rewarding positioning. Deep learning models are competitive in the bull regime: TimesNet reaches $+164.04\%$ and Informer $+105.12\%$, while MCap Hold achieves only $+21.13\%$. Single-agent models remain uniformly positive but modest (best: SA (CoT) $+27.20\%$), confirming that multi-agent coordination is primarily responsible for the performance gap.

\paragraph{Bear Market.} During the 15-week bear period, the ranking reverses: skill-augmented strategies sustain the largest losses (Debate (Skill) $-47.30\%$, Collaborative (Skill) $-43.87\%$), as momentum signals that drove bull-market gains amplify losses when trends reverse. In contrast, \ac{cot}-augmented strategies exhibit superior capital preservation: \emph{Hierarchical (CoT)} records the smallest loss at $-15.51\%$, with a maximum drawdown of only $-4.42\%$ and the lowest annualised volatility in the table ($31.25\%$). \emph{Collaborative (RAG)} also demonstrates resilience ($-22.13\%$, MDD $-10.42\%$), suggesting historical-analogue grounding moderates overreactions to short-term dislocations.

\subsection{Capability and Architecture Analysis}
\label{subsec:capability_analysis}

To isolate the contribution of each agent capability and \ac{mas} architecture, we examine performance patterns across the full $4 \times 3$ capability-by-architecture grid reported in \autoref{tab:performance}. Three patterns emerge.

First, \textbf{skill augmentation amplifies directional bets}: across all three architectures, the Skill variant achieves the highest cumulative return in the full period and the bull regime, but also incurs the highest volatility and drawdowns. Technical indicator signals are informative in trending markets but provide insufficient downside protection when trends reverse; the Skill capability is therefore best suited to bullish or momentum-driven environments.

Second, \textbf{\ac{cot} improves risk management}: \ac{cot}-augmented strategies consistently achieve lower volatility and shallower drawdowns than their zero-shot counterparts, particularly in the bear regime. The externalised reasoning chain appears to encourage more deliberate and conservative position sizing, a pattern consistent with findings in general financial reasoning tasks~\cite{Koa2024LearningModels}.

Third, \textbf{\ac{rag} reduces volatility at the cost of return}: \ac{rag} variants systematically exhibit lower annualised volatility than zero-shot variants across all three architectures. Collaborative (RAG) achieves the lowest annualised volatility of any strategy in the full period ($40.44\%$) and the bear regime ($40.08\%$). However, this conservatism limits upside capture: no \ac{rag} variant ranks in the top three for cumulative return in either the full period or the bull regime.

Across architectures, the hierarchical design provides the best risk-adjusted return in the full period (Hierarchical (Skill) SR $= +1.502$) and the strongest capital preservation in the bear market (Hierarchical (CoT) MDD $= -4.42\%$), suggesting that the supervisor-analyst structure effectively balances return generation with downside control. The debate architecture achieves the highest raw returns in bull conditions (Debate (Skill) $+290.96\%$) but is prone to amplified losses in adverse markets, reflecting the nature of adversarial argumentation: it surfaces conviction in positions, which is beneficial when those positions are correct and detrimental when they are not.

\subsection{Cross-Model Comparison}
\label{subsec:model_comparison}

\begin{figure}
  \centering
  \begin{subfigure}{\linewidth}
      \centering
      \includegraphics[width=\linewidth]{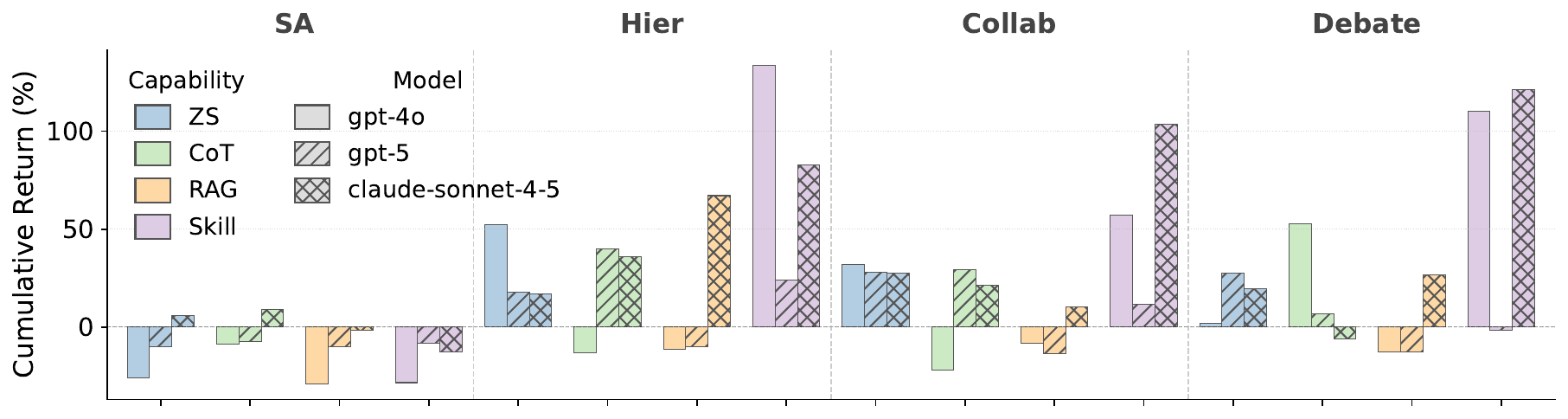}
      \subcaption{Cumulative Return.}
      \label{subfig:model_comparison_cum_ret}
  \end{subfigure}
  \hfill
  \begin{subfigure}{\linewidth}
      \centering
      \includegraphics[width=\linewidth]{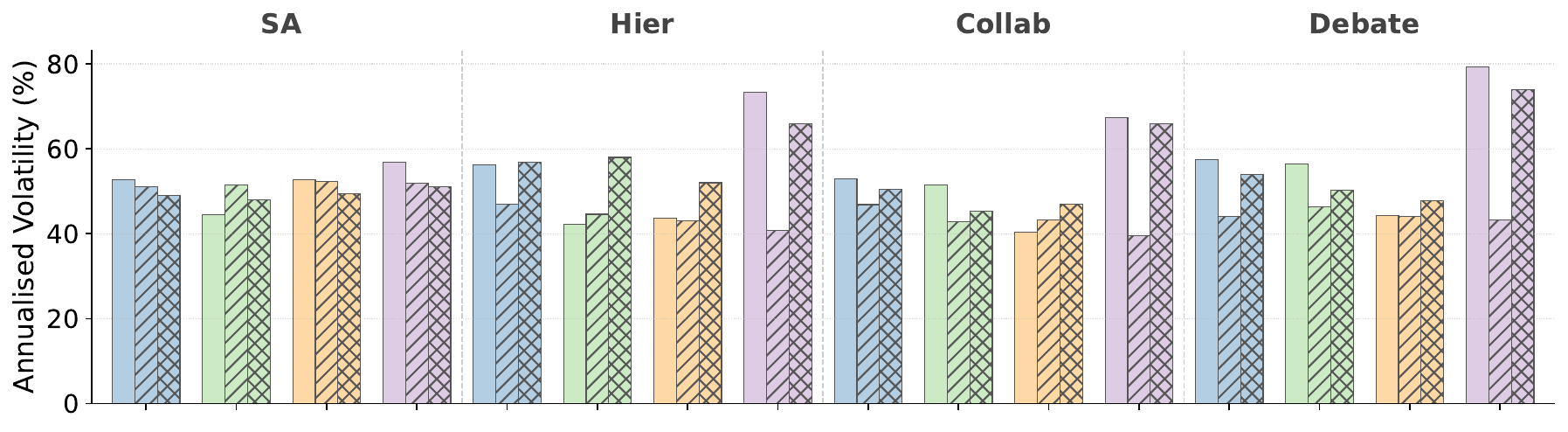}
      \subcaption{Annualized Volatility.}
      \label{subfig:model_comparison_ann_vol}
  \end{subfigure}
  \caption{Cross-model comparison of GPT-4o, GPT-5, and Claude Sonnet~4.5 across all 16 combinations.}
  \label{fig:model_comparison}
\end{figure}

\autoref{fig:model_comparison} compares GPT-4o, GPT-5, and Claude Sonnet~4.5 across all 16 architecture--capability combinations on cumulative return (\autoref{subfig:model_comparison_cum_ret}) and annualised volatility (\autoref{subfig:model_comparison_ann_vol}). Solid bars correspond to GPT-4o, hatched bars to GPT-5, and cross-hatched bars to Claude Sonnet~4.5; bar colour encodes the capability variant. Three observations emerge.

First, \textbf{Claude Sonnet~4.5 achieves the highest mean return while GPT-5 remains the most conservative backbone}. Averaged across all 16 combinations, Claude delivers a mean cumulative return of $+33.0\%$, ahead of GPT-4o ($+17.5\%$) and GPT-5 ($+6.9\%$), at similar average annualised volatility to GPT-4o ($54.1\%$ vs.\ $54.5\%$); GPT-5 is the least volatile ($45.8\%$). Claude's advantage is concentrated in skill-augmented \ac{mas} configurations: Debate (Skill) returns $+121.3\%$ and Collaborative (Skill) $+103.7\%$, each exceeding their GPT-4o equivalents ($+110.0\%$ and $+57.1\%$). GPT-4o nonetheless holds the single best result---Hierarchical (Skill) at $+133.5\%$---underscoring that model superiority is architecture-dependent rather than universal.

Second, \textbf{the three models exploit different capability configurations}. GPT-4o extracts the most alpha under skill augmentation, while GPT-5 benefits more from chain-of-thought prompting: Hierarchical (CoT) returns $+39.7\%$ under GPT-5 versus $-13.0\%$ under GPT-4o, and Collaborative (CoT) shows a comparable reversal ($+29.3\%$ vs.\ $-21.9\%$). Claude is the only model for which \ac{rag} augmentation delivers consistently strong \ac{mas} performance: Hierarchical (RAG) returns $+67.1\%$ under Claude versus near zero under both GPT-4o and GPT-5, suggesting that historical-analogue grounding interacts favourably with Claude's retrieval capabilities. Across architectures, the hierarchical design remains the strongest on a risk-adjusted basis under all three models.

Third, \textbf{\ac{mas} consistently outperforms the single-agent baseline for all three models}. Under GPT-4o, the average return across the 12 \ac{mas} configurations is $+31.0\%$ versus $-23.0\%$ for single-agent variants; under GPT-5, $+12.2\%$ vs.\ $-9.0\%$; under Claude, $+43.9\%$ against a near-breakeven single-agent average of $+0.1\%$, confirming that role specialisation and structured inter-agent communication provide consistent benefit regardless of backbone.

\subsection{Ablation Study}
\label{subsec:ablation}

\begin{table}[t]
\centering
\caption{Agent component ablation study (GPT-4o backbone). Each variant removes one component from the Hierarchical (ZS) reference system ($\dagger$). Deltas $(\Delta)$ relative to the reference are annotated with \inc{} (increase) and \dec{} (decrease).}
\label{tab:ablation}
\footnotesize
\setlength{\tabcolsep}{6pt}
\renewcommand{\arraystretch}{1.15}
\resizebox{\linewidth}{!}{
\begin{tabular}{l | rrrr}
\toprule
\multirow{2}{*}{\textbf{Strategy}} & \multicolumn{4}{c}{\textbf{Full Period} ($N=52$)} \\
\cmidrule(l){2-5}
& Cum\%\textcolor{red}{$\uparrow$} & Vol\%\textcolor{teal}{$\downarrow$} & SR\textcolor{red}{$\uparrow$} & Win\%\textcolor{red}{$\uparrow$} \\
\midrule
Hier.\ (ZS)$\dagger$ & $+52.19$ & $56.24$ & $+1.022$ & $61.5$ \\
\midrule
$-$ News Agent & $+51.31${\scriptsize \dec{0.88\%}} & $63.00${\scriptsize \inc{6.76\%}} & $+0.968${\scriptsize \dec{0.05}} & $57.7${\scriptsize \dec{3.80\%}} \\
$-$ Crypto Agent & $+9.62${\scriptsize \dec{42.57\%}} & $45.33${\scriptsize \dec{10.91\%}} & $+0.424${\scriptsize \dec{0.60}} & $50.0${\scriptsize \dec{11.50\%}} \\
$-$ Memory & $+40.72${\scriptsize \dec{11.47\%}} & $59.61${\scriptsize \inc{3.37\%}} & $+0.857${\scriptsize \dec{0.17}} & $55.8${\scriptsize \dec{5.70\%}} \\
\bottomrule
\end{tabular}
}
\end{table}

To quantify the contribution of each system component, we conduct a controlled ablation using GPT-4o as the backbone and \emph{Hierarchical (ZS)} as the reference, the simplest well-performing \ac{mas} configuration, and construct three variants, each removing exactly one component: the News Agent ($-$News Agent), the Crypto Agent ($-$Crypto Agent), and the rolling memory mechanism ($-$Memory). Absent components are replaced with zero-valued placeholder outputs so that the remaining agents are unaffected. \autoref{tab:ablation} reports all four metrics over the full 52-week period. Three findings emerge.

First, \textbf{the Crypto Agent is the most critical component}: its removal produces the largest degradation across all four metrics, reducing cumulative return by $42.57$ percentage points to just $+9.62\%$ and collapsing the Sharpe ratio from $+1.022$ to $+0.424$. The win rate falls to $50.0\%$, statistically indistinguishable from a coin flip, confirming that per-asset market signals are the primary driver of directional alpha.

Second, \textbf{memory provides meaningful continuity}: removing the rolling memory window costs $11.47$ percentage points in cumulative return ($+40.72\%$ versus $+52.19\%$), reduces the Sharpe ratio by $0.165$, and lowers the win rate by $5.70$ percentage points, reflecting the loss of cross-week trend-awareness that grounds week-to-week positioning.

Third, \textbf{the News Agent primarily modulates risk rather than return}: its removal reduces cumulative return by only $0.88$ percentage points but increases annualised volatility by $6.76$ percentage points and lowers the win rate by $3.80$ percentage points, indicating that news sentiment functions as a risk-dampening signal rather than a source of directional alpha.

\section{Conclusion}

We proposed an \ac{llm}-powered \ac{mas} framework for cryptocurrency portfolio management in which three modality-specialised agents operate under a $4 \times 3$ grid of capability and architecture configurations. The best-performing variant, \emph{Hierarchical (Skill)}, outperforms all single-agent, deep learning, and passive baselines, and a cross-model comparison with GPT-5 and Claude Sonnet~4.5 confirms that the \ac{mas} advantage is backbone-agnostic. Regime analysis reveals systematic architecture--capability trade-offs, while ablation identifies the Crypto Agent as the primary alpha driver and news sentiment and memory as risk modulators. All experiments use temperature $0.0$ to suppress sampling stochasticity; a full multi-seed sweep across the grid, three backbones, and 52 weekly ReAct rollouts is cost-prohibitive under commercial API pricing, and we leave seed-variance quantification to future work.

\bibliographystyle{IEEEtran}
\bibliography{references}
\appendix

\section{Agent System Instructions}
\label{apx:instructions}


\begin{instruc}[label={crypto_instruc},nameref={crypto_instruc}]{Crypto Agent.}
You are the Crypto Agent in a multi-agent cryptocurrency portfolio management system.

Your role is to analyse recent market data for a set of cryptocurrencies and produce a directional signal for each one.

For each asset you will receive the last $N$ days of: \textbf{close} (daily closing price in USD), \textbf{volume} (daily trading volume in USD), and \textbf{market\_cap} (daily market capitalisation in USD). Dates are in ascending order (oldest first, most recent last).

Produce a JSON array with one object per asset:
\begin{itemize}[leftmargin=10pt]
    \item \texttt{"symbol"} — ticker string
    \item \texttt{"signal"} — float in $[-1.0,\,1.0]$ \enspace ($-1=$ strong bearish,\; $+1=$ strong bullish)
    \item \texttt{"confidence"} — float in $[0.0,\,1.0]$
    \item \texttt{"rationale"} — one-sentence explanation
\end{itemize}
Return ONLY valid JSON, no prose outside the JSON.
\end{instruc}

\begin{instruc}[label={trading_instruc},nameref={trading_instruc}]{Trading Agent.}
You are the Trading Agent in a multi-agent cryptocurrency portfolio management system.

You receive: (1) Crypto Agent signals — technical scores for each asset (signal in $[-1,1]$, confidence in $[0,1]$); (2) News Agent signals — overall market sentiment and per-coin news signals; (3) current portfolio state — cash balance, per-asset holdings with unrealised P\&L, total value, and overall P\&L.

Your task is to produce executable trading actions for every asset in the universe.

\textbf{Action semantics:}
\begin{itemize}[leftmargin=10pt]
    \item Positive $(0,\,1]$: \textbf{BUY} — allocate that fraction of the post-sell cash pool to purchase the asset.
    \item Negative $[-1,\,0)$: \textbf{SELL} — liquidate that fraction of current holdings.
    \item Zero: \textbf{HOLD} — no change.
\end{itemize}

\textbf{Execution model:} All sell actions execute first; all buy actions then execute simultaneously from the resulting cash pool. If the sum of buy fractions exceeds 1.0, they are scaled down proportionally.

\textbf{Constraints:} Do not sell more than 100\% of a holding (action $\geq -1.0$). Avoid excessive concentration in any single asset. Be conservative when signals are conflicting or confidence is low. Use per-asset unrealised P\&L to inform profit-taking and loss-cutting decisions.

Produce a JSON array with one object per asset (include ALL assets):
\begin{itemize}[leftmargin=10pt]
    \item \texttt{"symbol"} — ticker string
    \item \texttt{"action"} — float in $[-1.0,\,1.0]$ \enspace (0.0 = hold)
    \item \texttt{"rationale"} — one-sentence justification referencing signals and/or P\&L
\end{itemize}
Return ONLY valid JSON, no prose outside the JSON.
\end{instruc}

\begin{instruc}[label={cot_instruc},nameref={cot_instruc}]{Chain-of-Thought Decorator (applied to all agents).}
Before giving your final answer, think step-by-step inside \texttt{<reasoning>...</reasoning>} tags. Your final structured JSON output must appear \emph{after} the closing \texttt{</reasoning>} tag.
\end{instruc}

\end{document}